# All-in-all-out magnetic domain wall conduction in pyrochlore iridate heterointerface

T. C. Fujita[1], M. Uchida[1,*], Y. Kozuka[1], W. Sano[1], A. Tsukazaki[1,2,3], T. Arima[4,5], M. Kawasaki[1,5]

**Affiliations**

[1]*Department of Applied Physics and Quantum-Phase Electronics Center (QPEC), University of Tokyo, Tokyo 113-8656, Japan*

[2]*Institute for Materials Research, Tohoku University, Sendai 980-8577, Japan*

[3]*PRESTO, Japan Science and Technology Agency (JST), Tokyo 102-0075, Japan*

[4]*Department of Advanced Materials Science, University of Tokyo, Kashiwa 277-8561, Japan*

[5]*RIKEN Center for Emergent Matter Science (CEMS), Wako 351-0198, Japan*

# ABSTRACT

Pyrochlore oxides possessing "all-in-all-out" spin ordering have attracted burgeoning interest as a rich ground of emergent states. This ordering has two distinct types of magnetic domains (all-in-all-out or all-out-all-in) with broken time-reversal symmetry, and a non-trivial metallic surface state has been theoretically demonstrated to appear at their domain wall. Here, we report on observation of this metallic conduction at the single all-in-all-out/all-out-all-in magnetic domain wall formed at the heterointerface of two pyrochlore iridates. By utilizing different magnetoresponses of them with different lanthanide ions, the domain wall is controllably inserted at the heterointerface, the surface state being detected as anomalous conduction enhancement with a ferroic hysteresis. Our establishment paves the way for further investigation and manipulation of this new type of surface transport.

Topological materials, involving topologically nontrivial band structure due to strong spin-orbit coupling (SOC), have been lately one of the central topics in the field of condensed matter physics. These materials are characterized by the edge state, which is uniquely determined by a type of symmetries inherent in the bulk state [1–10]. Recent theoretical studies have proposed that transition metal oxides such as iridates with heavy $5d$ electrons exhibit novel topological phases originating from an interplay of large SOC and electron correlation [11–14]. Engineering their band structure has been also suggested in the form of thin film or superlattice for realizing further emergent phases [15–22].

The pyrochlore oxide is well known as a fertile ground for various magnetic properties [23] such as spin ice in titanates [24,25] and large anomalous Hall effect induced by Berry phase in molybdates [26]. In the case of pyrochlore iridates ($Ln_2Ir_2O_7$; $Ln$: lanthanides), the system exhibits a metal-insulator transition accompanied with "all-in-all-out" spin ordering [13], where all four spins at the vertices in the tetrahedral cation site pointing inward or outward are alternatingly stacked along <111> direction [27,28]. In this antiferromagnetic ordering, there exist two distinct spin arrangements [all-in-all-out or all-out-all-in; Fig. 1(a)], which are interexchangeable with each other by the time-reversal operation. Hereafter we refer these two spin arrangements as "A domain" and "B domain" for simplicity. This antiferromagnetic ordering with broken time-reversal symmetry is one of the prerequisites for the emergence of the Weyl semimetal (WSM) phase in pyrochlore iridates [13].

In connection with the WSM phase, the metallic surface state has been theoretically demonstrated to appear at the domain wall between the A and B domains, even if the respective domains are not in the WSM phase [29–32]. This originates from the continuity of the low-energy bands which are classified by a symmetry of the wave functions near the Fermi level [Figs. 1(b) and 1(c)]. Actually, huge negative magnetoresistance ascribed to the magnetic domain wall conduction has been partly observed in $Nd_2Ir_2O_7$ (NIO) bulk crystals [33–35]. However, for bulk crystals containing randomly distributed magnetic domain walls in high density, it is highly difficult to investigate and control the domain wall conduction. Realizing the *single* magnetic domain wall at a well-defined surface or interface, as in the case of other topological materials, is inevitable for fostering the study of this new type of surface transport.

In this Rapid Communication, we study the metallic conduction emerging at a *single* all-in-all-out/all-out-all-in magnetic domain wall in the pyrochlore iridates. In order to realize this, we focus on pyrochlore iridate heterointerface composed of $Eu_2Ir_2O_7$ (EIO) and $Tb_2Ir_2O_7$ (TIO), where A or B domains can be independently controlled as "pinned layer" and "free layer", respectively [Fig. 1(d)]. When $Ln^{3+}$ has no magnetic moments such as $Eu^{3+}$ ($J = 0$), the magnetic domain structure of $Ir^{4+}$ is determined only by cooling magnetic field $B_{cool}$ and highly robust against sweeping magnetic field $B_{sweep}$ [36]. When $Ln^{3+}$ has magnetic moments such as $Tb^{3+}$ ($J = 6$), A or B domain can be switched each other by $B_{sweep}$ at some coercive field $B_c$ [27,34,35]. In this heterointerface, therefore, the magnetic domain wall is controllably inserted by the field control, B-EIO/A-TIO for $B_{sweep} > B_c$ after $B_{cool} < 0$ or A-EIO/B-TIO for $B_{sweep} < -B_c$ after $B_{cool} > 0$ [Fig. 1(d)].

The (111)-oriented pyrochlore iridate thin films and heterostructure were prepared on commercial YSZ (111) single crystal substrates by pulsed laser deposition [36]. We used a phase-mixed ceramics target with a prescribed ratio of Eu (Tb)/Ir = 1/3 fabricated by a hot-press method at 950 °C under 25 MPa pressure. The films were deposited at a substrate temperature of 500 °C under an atmosphere of 100 mTorr Ar gas containing 1% $O_2$. The KrF eximer laser ($\lambda$ =248 nm) was used for ablating the target, where the fluence and the frequency were 6 J/cm$^2$ and 10 Hz, respectively. The films were in an amorphous phase after the deposition, and pure pyrochlore phases appeared by annealing in an electrical muffle furnace in air at 1000 °C for 1.5 hours.

Magnetotransport measurements were carried out with conventional four-terminal method by using a liquid He cryostat equipped with a superconducting magnet (PPMS, Quantum Design Co.). Typical channel size was 500 × 500 μm$^2$, and current ($I$) and magnetic field ($B$) were applied parallel and perpendicular to the film surface, respectively. Here longitudinal conductance ($G_{xx}$) is obtained as inverse of the sheet resistance ($R_{xx}$); $G_{xx} = 1 / R_{xx}$ because contribution from the Hall resistance ($R_{xy}$) is negligibly small. For the field cooling measurement, cooling magnetic field ($B_{cool}$) was applied at 200 K perpendicular to the film surface and the sample was cooled to 10 K. Such measurement temperature below 20 K is important for switching the magnetic domain structure in TIO since the switching of Ir$^{4+}$ ordering is triggered by that of Tb$^{3+}$ whose ordering temperature is about 20 K [27,34,35]. The magnetotransport was measured as follows: the magnetic field was first set to $B_{cool}$, then was swept to −$B_{cool}$, and finally went back to the initial $B_{cool}$. $B_{cool}$ was ±9 T for EIO and heterostructure, and ±14 T for TIO, respectively.

A EIO (15 nm) / TIO (60 nm) heterostructure was epitaxially grown on an (111)-oriented Y-stabilized ZrO$_2$ (YSZ) single crystal substrate by combining pulsed laser deposition and solid phase epitaxy [See [36] for further detail]. X-ray diffraction (XRD) scan along [111] direction for heterostructure in Figs. 2(a) and 2(b) clearly shows separated peaks of EIO and TIO, indicating that both of the pyrochlore iridate layers are epitaxially formed. While the bottom TIO lattice is relaxed from YSZ, that of top EIO layer is strained by the bottom TIO as shown in the reciprocal space mapping in Fig. 2(c). A phase contrast image of the transmission electron microscopy (TEM) in Fig. 2(d) clearly shows the formation of the designed epitaxial heterostructure. We also performed energy dispersive x-ray spectrometry (EDX) in order to exclude a possibility that Eu and Tb atoms are intermixed each other. In spite of the solid phase epitaxy technique, interdiffusion of these two elements is fairly suppressed as shown in Fig. 2(f). An integrated EDX intensity profile in Fig. 2(e) also represents a very sharp interface within the resolution of measurement. All of these film characterizations ensure high quality of the heterostructure suitable for detecting the interfacial carrier transport.

Let us start by discussing transport property of the single layer films. The respective layers exhibit metal-insulator transition accompanied with the all-in-all-out magnetic ordering at Néel temperature $T_N$, as shown in Fig. 3(a). Magnetoresponse in the all-in-all-out spin ordering is generally described in the form of the third rank tensor [37]. Thus magnetoconductance (MC) includes the magnetic field ($B$)–linear term due to the canting of the $Ir^{4+}$ magnetic moments in the all-in-all-out ordering, the sign of which is opposite depending on the domain structure (A or B) [36]. Another distinct point is that, as mentioned above, MC is enhanced when the magnetic domain walls are formed in the process of the domain switching [33–35]. This response is completely opposite to the conventional magnetic materials, where conductance is suppressed due to domain wall scattering when the magnetic domains are mixed.

$B_{sweep}$ dependences of MC in the EIO and TIO films for various $B_{cool}$ are shown in Figs. 3(b) and 3(c), which are basically consistent with the previous reports [33-36]. In short, the domain of EIO is uniquely determined by $B_{cool}$, while that of TIO is switchable by $B_{sweep}$. For EIO [Fig. 3(b)], MC has finite $B$-linear component after field cooling ($B_{cool} = +9$ or $-9$ T, FC), reflecting that the magnetic domain structure is fixed to A or B single domain by $B_{cool}$. After zero-field cooling ($B_{cool} = 0$ T, ZFC), on the other hand, the $B$-linear component is cancelled out due to mixture of the domains. It is worth mentioning that MC does not show any hysteresis because the magnetic domain structure is quite robust against $B_{sweep}$. In the case of TIO [Fig. 3(c)], on the other hand, MC shows hysteresis originating from the domain switching at $B_c \sim 8$ T, triggered by the switching of all-in-all-out ordering in $Tb^{3+}$ lattice formed below about 20 K [27,34,35]. Starting from $B_{sweep}$ well above $B_c$, where the magnetic domain is regulated as A domain (colored by red), $G_{xx}$ reduces while $B_{sweep}$ approaching 0 T. Then $B_{sweep}$ being swept to negative, where the magnetic domain is gradually switched to B domain and the A and B domains are intermixed, MC gets enhanced compared with the reverse sweep. Finally MC reaches its top at $-B_c$ and then suddenly drops to finish the entire switching to the B domain (colored by blue). The same process occurs when $B_{sweep}$ increases again and the magnetic domain is switched back from B to A.

We then move on to discuss magnetotransport of heterointerface between the two layers. We note again that the measured conductance is a summation of the contributions from EIO, TIO, and interface of them due to the measurement configuration of the heterostructure, as written as

$$G_{xx} = G_{xx}^{EIO} + G_{xx}^{TIO} + G_{int}. \tag{1}$$

As shown in Fig. 4(b), for ZFC, symmetric MC is observed in the EIO/TIO heterostructure. A similar dip-like structure appears around $B_{sweep} = 6$ T, with a hysteresis in the same direction as in TIO single layer film [Fig. 3(c)]. This is a clear evidence that the domain switching in the TIO layer occurs as in the case of the TIO single layer film, which is also verified by the minor-loop magnetotransport measurement (See Supplemental Material [38]). In the case of FC, $G_{xx}$ exhibits overall gradients with opposite sign for $B_{cool} = +9$ or $-9$ T [Figs. 4(a) and 4(c)]. The gradient is attributed to the $B$-linear MC in the EIO layer, whose magnetic domain is aligned to the single A (or B) domain depending on positive (or negative) $B_{cool}$ as in the case of the EIO single layer film [Fig. 3(b)].

Additional conductance ($G_{int}$), indicated by orange bars in Figs. 4(a) and 4(c), appears as a ferroic loop between the sweeps upward and downward for $-B_c < B_{sweep} < B_c$. This additional conductance emerges only when the domain structures are opposite between the TIO and EIO layers, suggesting that the theoretically predicted metallic surface state is materialized at the single magnetic domain wall induced at the heterointerface. The contribution, the sum of *B*-linear component in the EIO layer and the additional ferroic hysteresis at the heterointerface, can be extracted by subtracting $G_{xx}$ after ZFC [Fig. 4(b)] from those after FCs [Figs. 4(a) and 4(c)] as shown in Figs. 4(d) and 4(e), respectively, because any other contributions are commonly contained in ZFC. They can be further divided into the *B*-linear and ferroic terms, as shown in Fig. 4(f) (See Supplemental Material [38] for further information about deducing process). The ferroic hysteresis, which has opposite loop direction for the opposite cooling field, appears reflecting creation and annihilation of the single conduction path at the A/B magnetic domain wall. Here, we have to emphasize again that this hysteresis is purely due to the effect of heterointerface, because such additional conduction is not observed for single layers of EIO or TIO. The hysteresis height is almost independent of the magnetic field, which evidences that the hysteresis does not originate from the spin canting in disordered spin structure at the heterointerface. The height of the hysteresis, corresponding to the conductance at the interface ($G_{int}$), is about 0.4 µS., $G_{int}$ accounts for 0.2% of the total conductance, this ratio is about 20% for the previously reported bulk NIO. This is quite reasonable assuming that polycrystalline NIO contains multiple domain walls with a domain size of 50-100 µm.

For further investigation of the interfacial conductance, temperature dependence of $G_{int}$ [Fig. 5(a)] was taken with the following process. First, the sample was cooled down to 10 K with field cooling ($B_{cool}$ = +9 T), and at 10 K, the field was swept to 0 T. At this point, both the EIO and TIO layers forms the A domain. Then the $G_{xx}$ was measured from 2 K to 200 K under zero field (i). Second, the sample was also cooled down to 10 K with field cooling ($B_{cool}$ = +9 T), and at 10 K, the field was swept to −9 T and then returned to 0 T. At this point TIO is switched to the B domain, while EIO remains the A domain. Then the $G_{xx}$ was measured from 2 K to 200 K under zero field (ii). Comparing the case (i) with (ii), the magnetic domain wall is introduced only in the case (ii), as illustrated in Fig. 4(e), and thus the temperature dependence of $G_{int}$ is derived by subtracting (i) $G_{xx}$ from (ii) $G_{xx}$.

The obtained $G_{int}$ emerges below about 110 K, which is in good agreement with the all-in-all-out ordering temperature $T_N$ of the heterostructure [Fig. 5(b)]. This finding also confirms that $G_{int}$ originates from the surface state between the A and B domains. Decrease in $G_{int}$ below 30 K may be attributed to the electron correlation effect of the $5d$ electron bands [39], or the weak localization as generally expected in the two dimensional case. While the bulk conduction accounting for most of $G_{xx}$ is highly suppressed with decreasing temperature below $T_N \sim 105$ K, the interfacial conduction $G_{int}$ keeps increasing down to 30 K. This temperature dependence conclusively demonstrates that the interface conduction at the all-in-all-out magnetic domain wall is indeed metallic.

To summarize, we have fabricated pyrochlore iridate heterostructure in order to detect metallic conduction expected at the single all-in-all-out/all-out-all-in magnetic domain wall. We have demonstrated that the additional metallic conductance emerges only when the domain wall is inserted at the heterointerface by the field control. The single magnetic domain wall at the heterointerface is controllably induced, and it remains within a wide range of sweep magnetic field, different from ones in bulk crystals. Further investigation would verify topological nature of the magnetic domain wall state by measuring such as anomalous Hall effect or non-local transport properties on this single domain wall. Our establishment will promote future exploration of surface transport in the new type of topological phases in oxides.

## ACKNOWLEDGMENTS

We thank J. Fujioka, K. Ueda, Y. Tokura, N. Nagaosa, Y. Yamaji, M. Imada, M. Udagawa, and Y. Motome for helpful discussions. This work was partly supported by Grant-in-Aids for Scientific Research (S) No. 24226002 and No. 24224010, by JSPS Fellowship No. 26·10112 (TCF), and by Challenging Exploratory Research No. 26610098 (MU) from MEXT, Japan as well as by Asahi Glass Foundation (YK).
10

**FIGURES**

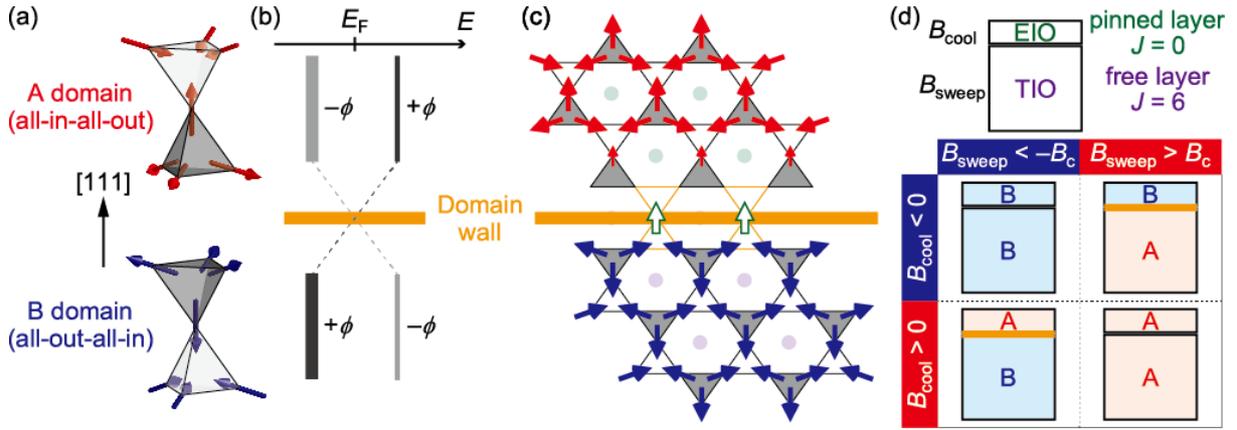

**FIG. 1 Fujita *et al*.,**

FIG. 1. (Color online) (a) Two distinct all-in-all-out spin structures, named as A (all-in-all-out) domain and B (all-out-all-in) domain, on pyrochlore lattice. (b), (c) Their eigenstates near the Fermi level ($E_F$) have different phase factors ($\pm\phi$) and they are switched by time-reversal operation, leading to the surface state at the domain wall accompanied with ferromagnetic moment indicated by open green arrows. (d) Cross sectional schematics of the heterostructure composed of $Eu_2Ir_2O_7$ (EIO) with $J = 0$ nonmagnetic $Eu^{3+}$ and $Tb_2Ir_2O_7$ (TIO) with $J = 6$ Ising-like $Tb^{3+}$, which are used as "pinned layer" and "free layer" and controlled by $B_{cool}$ and $B_{sweep}$, respectively. Their magnetic domain structures are changed depending on $B_{cool}$ and $B_{sweep}$. Domain wall conduction path (denoted by yellow lines) emerges when the domain structure is opposite in EIO and TIO layers.

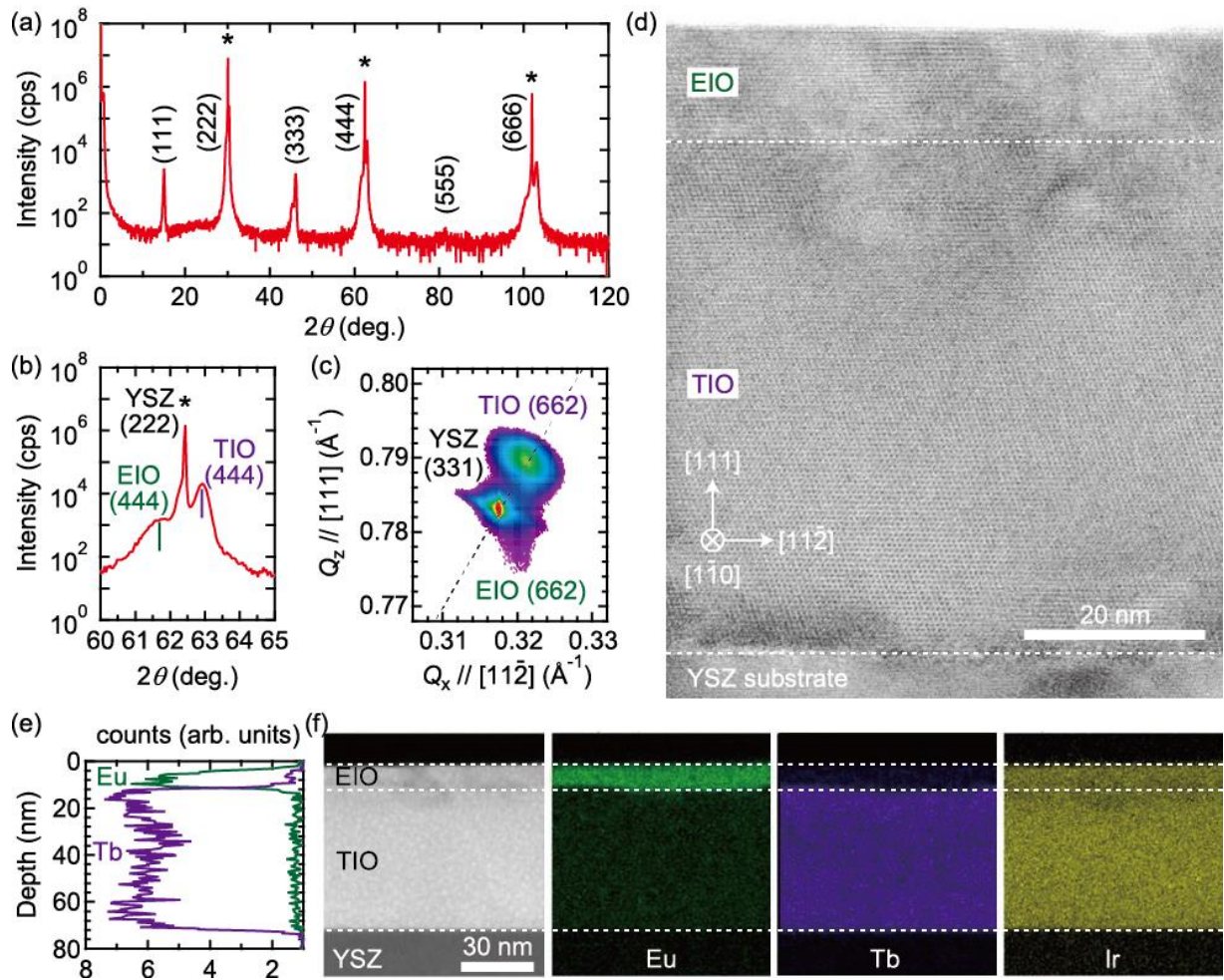

**FIG. 2 Fujita *et al*.,**

FIG. 2. (Color online) (a) $\theta$-$2\theta$ scan of x-ray diffraction. Peaks from the YSZ substrate are marked with asterisks. Magnified view around YSZ (222) is shown in (b). (c) Reciprocal space mapping around YSZ (331) peak. Ideal pyrochlore lattice with cubic symmetry follows the dashed line. (d) Phase contrast image of high-resolution TEM of the EIO / TIO heterostructure on YSZ (111) substrate. (e) Integrated intensity along [111]-direction of the energy dispersive x-ray spectrometry (EDX) for Eu and Tb. (f) Magnified HHADF-STEM image and cross sectional EDX mappings for Eu, Tb and Ir elements.

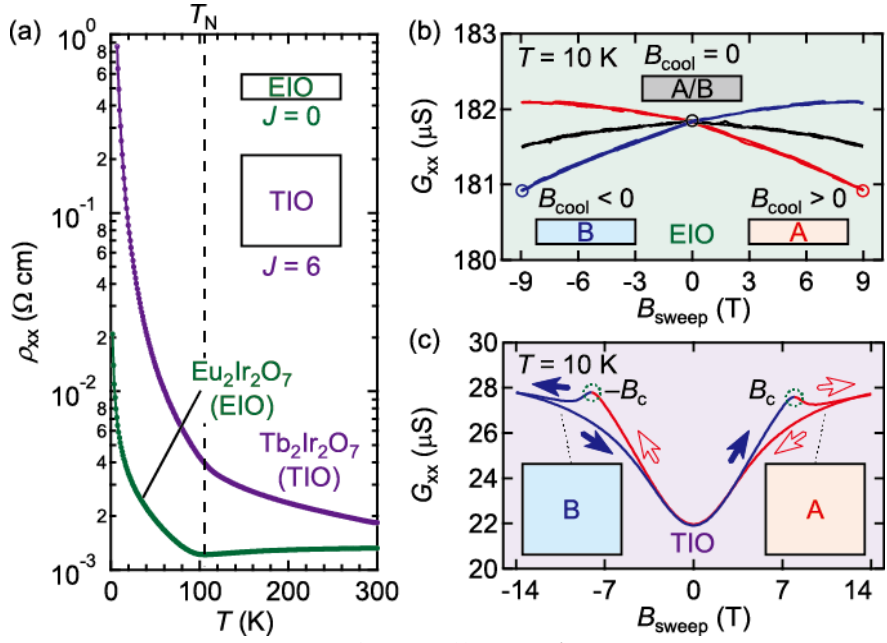

**FIG. 3 Fujita et al.,**

FIG. 3. (Color online) (a) Temperature dependence of the longitudinal resistivity ($\rho_{xx}$) for 15nm $Eu_2Ir_2O_7$ (EIO) and 60 nm $Tb_2Ir_2O_7$ (TIO) thin films. Magnetic transition temperature $T_N$ is indicated by dashed line. (b), (c) Sweeping magnetic field ($B_{sweep}$) dependence of longitudinal conductance $G_{xx}$ for EIO (b) and TIO (c) measured at 10 K. Magnetic domain structure of EIO is determined by cooling field ($B_{cool}$), whereas that of TIO can be switched by $B_{sweep}$. The direction of hysteresis originating from domain switching is shown by arrows, where open (filled) red (blue) one corresponds to the domain structure for A (B) domain.

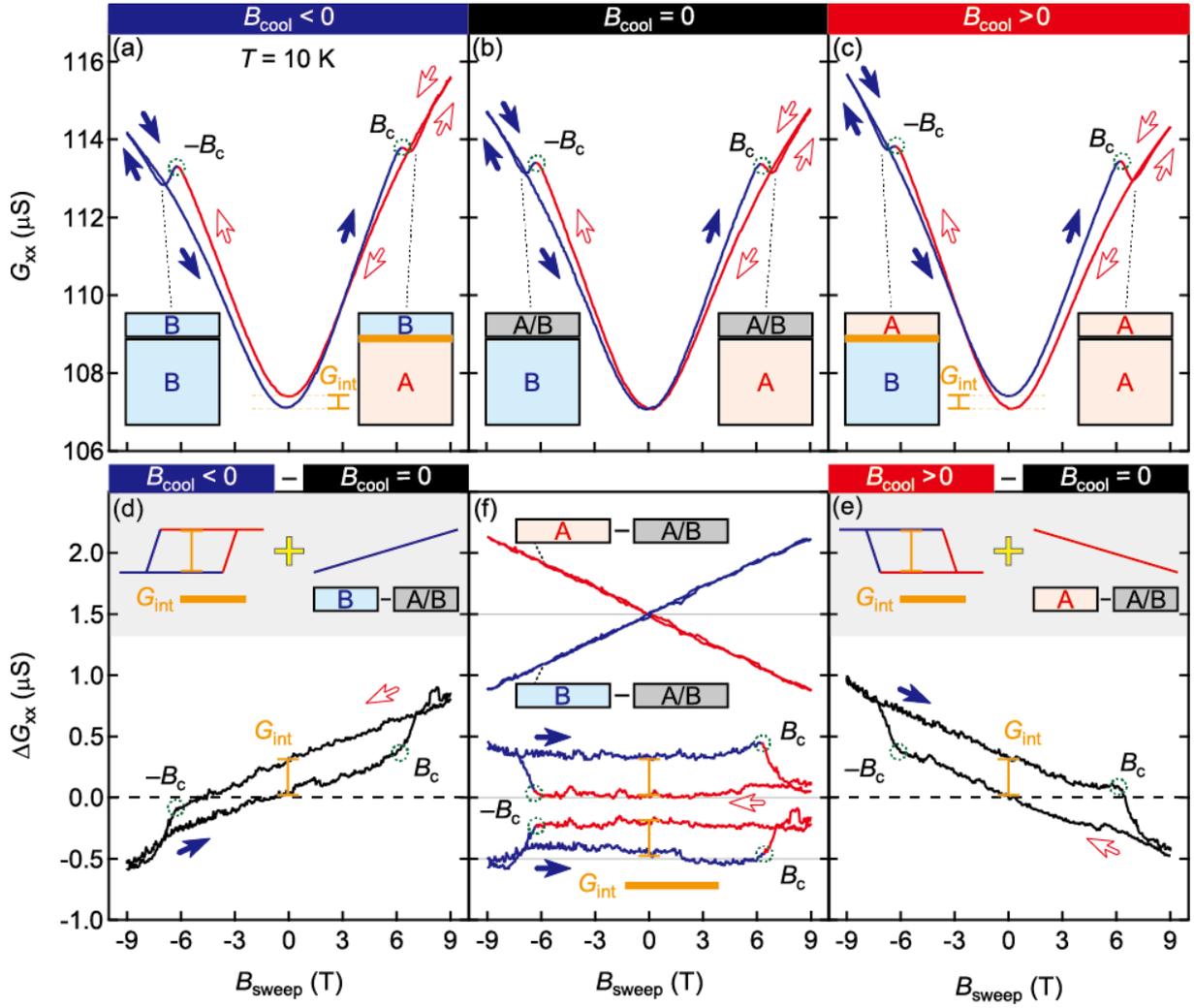

**FIG. 4 Fujita et al.,**

FIG. 4. (Color online) (a)-(c) Sweeping magnetic field ($B_{sweep}$) dependence of longitudinal conductance $G_{xx}$ for EIO / TIO heterostructure measured at 10 K after various cooling field ($B_{cool}$) conditions. The line colors indicate the magnetic domain structure in TIO layer (red (blue) for A (B) domain). The direction of hysteresis is shown by arrows, whose colors also indicate the magnetic domain structure in TIO layer. (d), (e) $B_{sweep}$ dependence of the subtracted conductance defined by $G_{xx}$ (±9 T FC) − $G_{xx}$ (ZFC) for the heterostructure. These results correspond to the summation of linear magnetoconductance in EIO layer and additional conductance at the interface ($G_{int}$). (f) $B_{sweep}$ dependence of the linear magnetoconductance (upper half) and $G_{int}$ (lower half) resolved from (d) and (e).

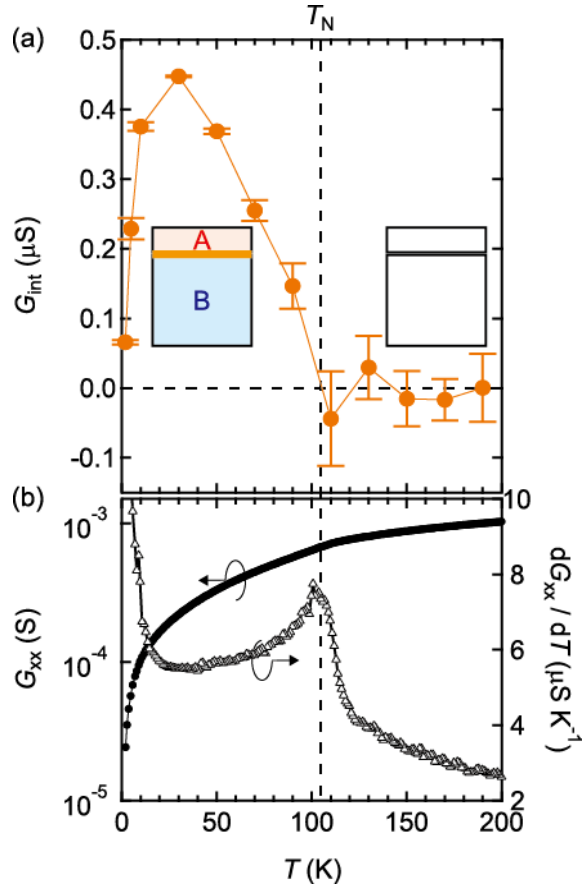

**FIG. 5 Fujita *et al.*,**

FIG. 5. (Color online) (a), (b) Temperature dependence of the interfacial conductance ($G_{int}$), compared with that of the total longitudinal conductance $G_{xx}$ of EIO / TIO heterostructure. Magnetic transition temperature $T_N$ is estimated to be about 105 K from the peak in the temperature derivative.

# Supplementary Material
# All-in-all-out magnetic domain wall conduction in pyrochlore iridate heterointerface

T. C. Fujita[1], M. Uchida[1,*], Y. Kozuka[1], W. Sano[1], A. Tsukazaki[1,2,3], T. Arima[4,5], M. Kawasaki[1,5]

**Affiliations**

[1]*Department of Applied Physics and Quantum-Phase Electronics Center (QPEC), University of Tokyo, Tokyo 113-8656, Japan*

[2]*Institute for Materials Research, Tohoku University, Sendai 980-8577, Japan*

[3]*PRESTO, Japan Science and Technology Agency (JST), Tokyo 102-0075, Japan*

[4]*Department of Advanced Materials Science, University of Tokyo, Kashiwa 277-8561, Japan*

[5]*RIKEN Center for Emergent Matter Science (CEMS), Wako 351-0198, Japan*

## Minor loop magnetotransport measurement

Figure S1 shows longitudinal conductance ($G_{xx}$) for the Eu$_2$Ir$_2$O$_7$ (EIO) / Tb$_2$Ir$_2$O$_7$ (TIO) heterostructure as a function of magnetic field ($B_{sweep}$) at 10 K after +9 T field cooling. For measuring the minor loops of TIO layer, magnetic field was swept as following sequence. First, the sample was cooled from 200 K under $B_{sweep}$ = +9 T. Then, at 10 K magnetic field was swept from +9 T to a certain field ($B_r$), which we chose $-6$ T [Fig. 5(a)], $-6.5$ T [Fig. 5(b)], $-7.2$ T [Fig. 5(c)], and $-9$ T [Fig. 5(d)], and swept back to +9 T. After the sweep for each $B_r$, the sample was heated up to 200 K again for the next measurement for different $B_r$. When $B_r = -6$ T, $G_{xx}$ does not show any hysteresis, indicating there was negligible domain reversal. In case of $B_r = -6.5$ T, it shows some finite but smaller hysteresis, indicating insufficient domain switching. On the other hand with $B_r = -7.2$ T, hysteresis appears with almost the same order of magnitude as in the case of full-loop. It is worth mentioning, however, that even in this case of $B_r = -7.2$ T, $G_{xx}$ does not show the full hysteresis as in $B_r = -9$ T, which is distinguished by the cross of upward sweep and down ward sweeps at around $B_{sweep} = -4$ T in Fig S1 (d). This indicates that some part of the magnetic domain is not reversed in this $B_r = -7.2$ T sequence. From these observations we conclude that the characteristic hysteresis shown in the $G_{xx}$ originates from the magnetic domain switching in TIO layer, and its coercive field for complete switching is around 7.5 T.

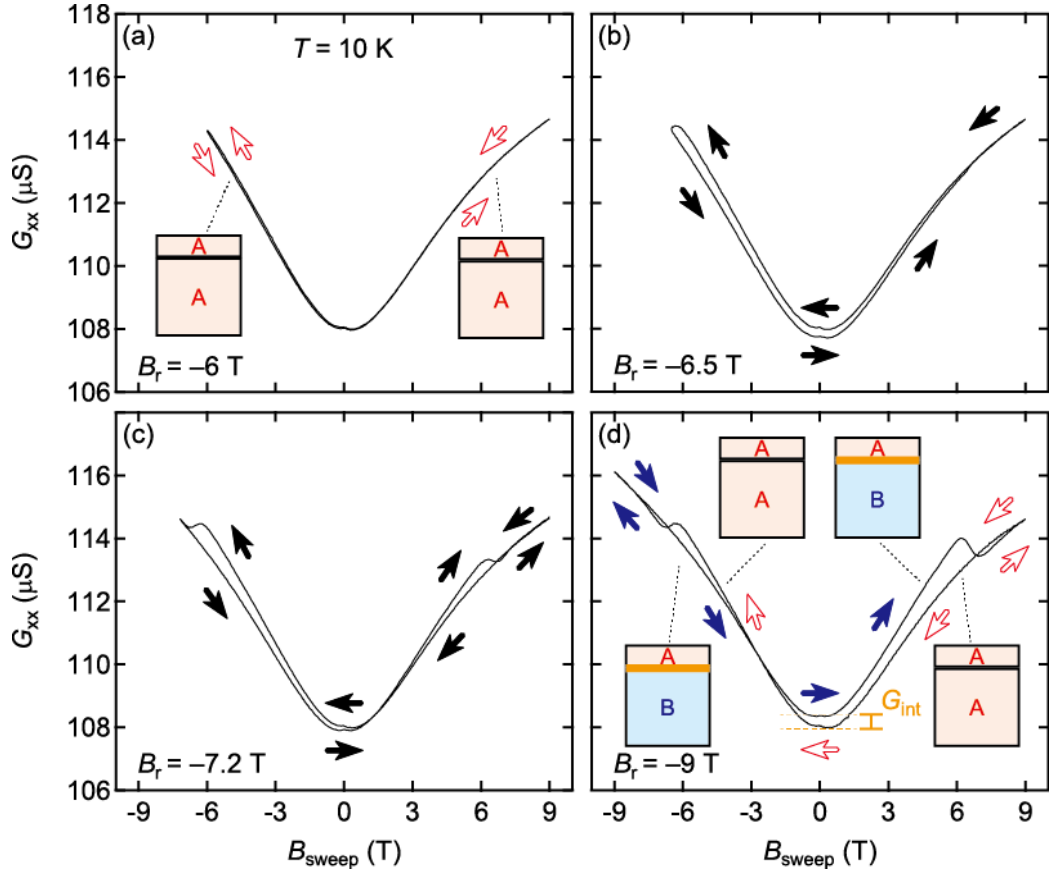

FIG. S1. Minor loop measurements for the EIO / TIO heterostructure. Longitudinal conductance ($G_{xx}$) for the Eu$_2$Ir$_2$O$_7$ (EIO) / Tb$_2$Ir$_2$O$_7$ (TIO) heterostructure as a function of magnetic field ($B_{sweep}$) at 10 K after +9 T field cooling. The magnetic field is swept as +9 T → $B_r$ → +9 T, where $B_r$ is the turning back point of the magnetic field. As for $B_r$, we chose −6 T (a), −6.5 T (b), −7.2 T (c), and −9 T (d).

**Calculation process for deducing interfacial conductance**

Magnetoconductance in pyrochlore iridates with all-in-all-out magnetic ordering generally has linear-term [37], and it is described as

$$G_{xx} = \alpha B + \sum_n b_n B^{2n} (+G_{DW}) \ (n = 0, 1, 2, …). \tag{S1}$$

Here, $B$ is external magnetic field, $G_{DW}$ is a domain wall conductance in single layer film, and the coefficient $\alpha$ is negative (positive) for A (B) domain.

At first, we start from the case without $G_{DW}$ in order to simplify the discussion. The magnetic domain in EIO is determined only by $B_{cool}$, and robust against $B_{sweep}$ at low temperature once it is formed [36]. Therefore, its magnetoconductance shows gradient $\alpha$ corresponding to the magnetic domain structure, namely $B_{cool}$. However, the domain in TIO is switched at certain magnetic field, where the sign of $\alpha$ is also inverted, and thus its magnetoconductance is seemingly even. In other words, the coercive field of $Ir^{4+}$ in EIO is much larger than that in TIO due to the absence of magnetic moment of $Eu^{3+}$. Therefore, if we were able to apply magnetic field huge enough to switch the magnetic domain in EIO, we would observe even magneto conductance as in the case of TIO.

Next, we consider the contribution from domain wall conductance in single layer film, namely $G_{DW}$. In EIO, magnetic domain is aligned to A (B) single domain with positive (negative) $B_{cool}$, and thus $G_{DW}$ should disappear. On the other hand, the magnetic domain after ZFC is a mixture of A and B domains, and $G_{DW}$ could appear. In TIO, the magnetic domain can be switched by $B_{sweep}$, and thus $G_{DW}$ is independent of $B_{cool}$.

Finally, we argue the conduction of heterointerface, $G_{int}$. It is supposed that $G_{int}$ emerges only when magnetic domains in TIO and EIO layers are different with each other. In our present work, we can switch only magnetic domain in TIO, and thus time-reversal symmetry is broken, leading to odd (ferroic) field dependence of $G_{int}$ [FIG. S2(a)]. If we applied sufficient magnetic field to invert magnetic domain in EIO, time-reversal symmetry would be conserved, and $G_\text{int}$ would show even field dependence [FIG. S2(b)].

Taking into account all the discussions above, the total conductance of the heterostructure after ZFC and ±9 T FCs are written as followings:

$$G_\text{xx}(\text{ZFC}) = G_\text{xx}^\text{EIO} + G_\text{xx}^\text{TIO} + G_\text{int} = (\sum_n b_n B^{2n} + G_\text{DW}^\text{EIO}) + (\alpha B + \sum_n b_n B^{2n} + G_\text{DW}^\text{TIO}). \quad (S2)$$

$$G_\text{xx}(\pm 9\,\text{T FC}) = G_\text{xx}^\text{EIO} + G_\text{xx}^\text{TIO} + G_\text{int} = (\mp \alpha B + \sum_n b_n B^{2n}) + (\alpha B + \sum_n b_n B^{2n} + G_\text{DW}^\text{TIO}) + G_\text{int}. \quad (S3)$$

Consequently, $\Delta G_\text{xx}$ shown in FIGs. 4(d) and 4(e) can be deduced as

$$\Delta G_\text{xx}(\pm 9\,\text{T FC}) \equiv G_\text{xx}(\pm 9\,\text{T FC}) - G_\text{xx}(\text{ZFC}) = (\mp \alpha B - G_\text{DW}^\text{EIO}) + G_\text{int}. \quad (S4)$$

$\Delta G_\text{xx}$ can be further divided into the $B$-linear ($\mp \alpha B - G_\text{DW}^\text{EIO}$) and ferroic ($G_\text{int}$) terms, which are shown in upper and lower half of FIG. 4(f), respectively.

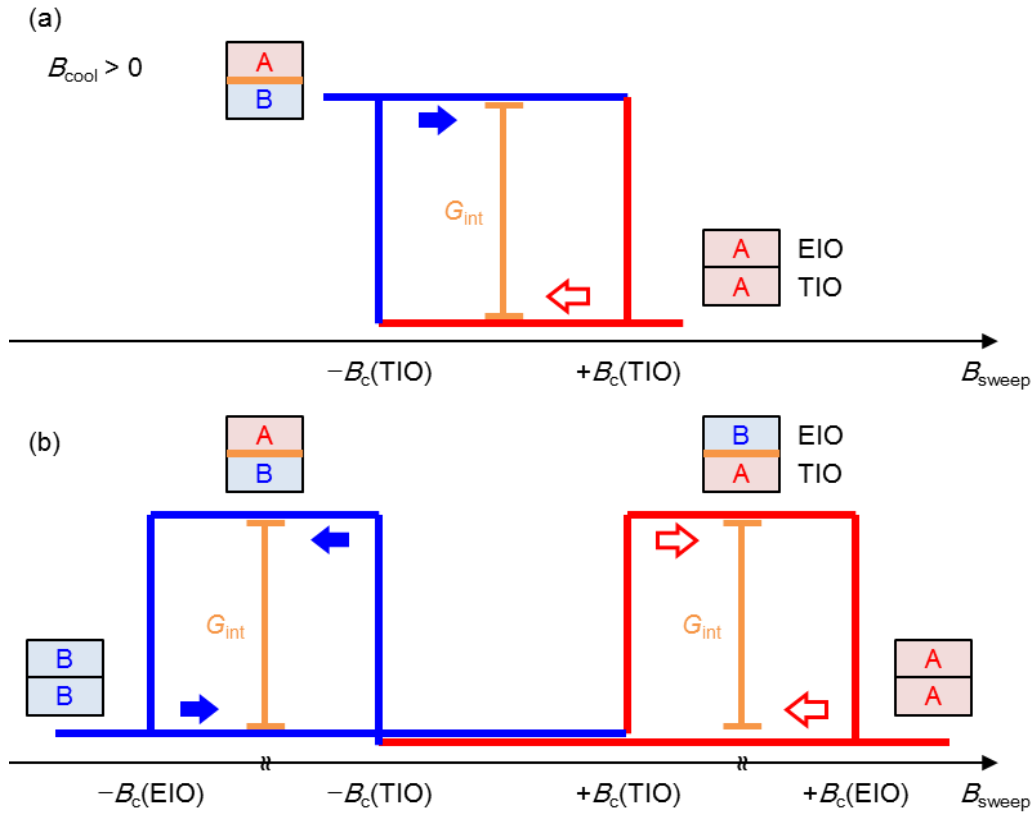

FIG. S2 Schematics of sweep magnetic field ($B_\text{sweep}$) dependence of $G_\text{int}$ within feasible (a) and unfeasible (b) ranges of magnetic field, respectively. Here, $B_\text{c}(\text{EIO})$ and $B_\text{c}(\text{TIO})$ denote coercive fields of EIO and TIO layers, respectively.